\documentclass[aps,prb,reprint,twocolumn,superscriptaddress,amsmath, amssymb]{revtex4-2}

\usepackage{graphicx}
\usepackage{txfonts}
\usepackage{orcidlink}
\usepackage{booktabs}

\hypersetup{
    colorlinks=true,
    allcolors=blue
}

\begin{document}

\title{Ab initio electronic conductivity of Fe-bearing post-perovskite}

\author{Yihang Peng\,\orcidlink{0000-0003-4404-8973}}
\email{yhpeng@princeton.edu}
\affiliation{Department of Geosciences, Princeton University, Princeton, NJ, USA}

\author{Yupei Zhang\,\orcidlink{0009-0000-5835-4619}}
\affiliation{Department of Physics, Temple University, Philadelphia, PA, USA}

\author{Shuai Zhang\,\orcidlink{0000-0001-9503-4964}}
\affiliation{Laboratory for Laser Energetics, University of Rochester, Rochester, NY, USA}

\author{Chenxing Luo\,\orcidlink{0000-0003-4116-6851}}
\affiliation{Department of Geosciences, Princeton University, Princeton, NJ, USA}
\affiliation{Department of Applied Physics and Applied Mathematics, Columbia University, New York, NY, USA}

\author{Donghao Zheng\,\orcidlink{0009-0003-0497-4961}}
\affiliation{Department of Geosciences, Princeton University, Princeton, NJ, USA}

\author{Nelson Naveas\,\orcidlink{0000-0003-0949-277X}}
\affiliation{
Departamento de Física Aplicada and Instituto Universitario de Ciencia de Materiales “Nicolás Cabrera” (INC), Universidad Autonoma de Madrid, Campus de Cantoblanco, Madrid, Spain}

\author{Xifan Wu\,\orcidlink{0000-0001-6561-8942}}
\affiliation{Department of Physics, Temple University, Philadelphia, PA, USA}

\author{Jie Deng\,\orcidlink{0000-0001-5441-2797}}
\email{jie.deng@princeton.edu}
\affiliation{Department of Geosciences, Princeton University, Princeton, NJ, USA}


% \date{\today}

\begin{abstract}
The electrical conductivity of high-pressure silicates profoundly influences the interior dynamics of rocky planets. Employing the Kubo-Greenwood formalism, we perform \textit{ab initio} calculations of electronic conductivity in Fe-bearing post-perovskite under super-Earth mantle conditions, up to 4000 K and 500 GPa. Electronic structures are obtained via many-body perturbation theory, incorporating dynamical screening and correlations among localized Fe-$3d$ orbitals. In contrast to (Fe,Mg)O, for which metallization has been reported at comparable conditions, our results indicate that post-perovskite with Earth-like Fe contents is unlikely to metallize in super-Earth mantles via band-gap closure, yielding negligible low-frequency conductivity. Any substantial conductivity would require non-electronic mechanisms, such as thermally activated small-polaron hopping, which fall beyond the scope of band conduction.
\end{abstract}

\maketitle


\textit{Introduction}---Understanding the dynamics and evolution of planets requires detailed knowledge of the transport properties of planet-forming materials. Electrical conductivity is a fundamental transport property that governs how planets respond to electromagnetic processes, whether generated internally by a planetary dynamo \cite{banks_geomagnetic_1969, buffett_constraints_1992, vilim_effect_2013} or externally through star-planet magnetic interactions \cite{kislyakova_magma_2017, peng_induction_2025}. Recent studies suggest that silicate liquid \cite{stixrude_silicate_2020} and (Fe,Mg)O \cite{ohta_experimental_2012, ohta_highly_2014, holmstrom_electronic_2018} may become metalized due to the closure of band gap under Earth's lowermost mantle conditions. However, the electrical conductivity of the dominant mantle mineral at corresponding and higher pressures ($\gtrsim$130~GPa \cite{murakami_post-perovskite_2004}), (Mg,Fe)SiO$_3$ post-perovskite (pPv), remain poorly understood. Since pPv is one of the most abundant phases in super-Earths (i.e., rocky exoplanets with masses ranging from 1--10 Earth masses) \cite{wagner_rocky_2012}, a comprehensive understanding of its conductivity is not only important for constraining the structure of Earth's deep mantle, but also essential for modeling the interior dynamics and potential habitability of super-Earths.

Accurately determining the electrical conductivity under the extreme conditions where pPv is stable remains experimentally challenging. In this context, theoretical calculations play an indispensable role in addressing this issue. However, first-principles methods encounter significant challenges in modeling the electronic conductivity of Fe-bearing silicates. These difficulties stem from the strongly correlated nature of Fe-$3d$ electrons, which conventional density functional theory (DFT) fails to capture adequately. Furthermore, the accuracy of such calculations is compromised by severe self-interaction errors and the absence of derivative discontinuities inherent in local and semi-local exchange-correlation functionals \cite{cohen_insights_2008}. Consequently, these limitations substantially undermine the predictive capability of standard DFT for describing the electronic structure and conductivity of Fe-bearing post-perovskite under the extreme conditions prevalent in planetary mantles. Although the DFT+$U$ method offers a straightforward and computationally efficient correction for localized $3d$ electrons in perovskites \cite{cococcioni_accurate_2010, wang_electronic_2017, wang_convert_2022}, it remains a static, localized mean-field approximation that retains significant self-interaction errors. Furthermore, the efficacy of this semi-empirical approach is highly sensitive to the Hubbard $U$ parameter, which varies with crystal structure, valence, and spin state \cite{cococcioni_linear_2005, hsu_spin-state_2010, floris_hubbard-corrected_2020}. On the other hand, hybrid DFT approaches \cite{heyd_hybrid_2003, paier_screened_2006} mitigate self-interaction errors and the lack of derivative discontinuities by incorporating a fraction of non-local exact exchange. Overall, hybrid DFT provides a more accurate description of transition metal-bearing materials, with improvements comparable to those from DFT+$U$ in many aspects, despite not treating correlated $d$-electrons in the same localized manner \cite{tran_hybrid_2006, johannes_hole_2012, zhang_re-evaluation_2019}. However, both the DFT+$U$ and hybrid functional methods neglect dynamical screening effects, leading to notable discrepancies between calculated and experimental spectra \cite{rodl_quasiparticle_2009, nohara_electronic_2009, lany_band-structure_2013}. The $GW$ approximation, an electronic structure method within many-body perturbation theory, incorporates dynamical screening effects while significantly mitigating self-interaction errors. This framework computes accurate quasiparticle (QP) energies, making it a powerful and widely used tool for predicting electronic structure in various materials in weakly to moderately correlated systems \cite{faleev_all-electron_2004, rinke_combining_2005, rodl_quasiparticle_2009, nohara_electronic_2009, lany_band-structure_2013}. However, this method has rarely been applied to geophysically relevant materials under extreme conditions. The compositional complexity of planet-forming materials and the finite-size effects inherent to conductivity calculations necessitate large systems \cite{stixrude_silicate_2020}; achieving accurate high-temperature conductivities further requires statistical averaging over many molecular dynamics (MD) snapshots \cite{ghosh_melting_2024}. These constraints render one-shot $G_0W_0$ the only practical option, but the sensitivity of its results to the mean-field starting point demands careful selection of the input wavefunctions \cite{liao_testing_2011}

In this study, we explore the electronic conductivity of pPv by combining multiple advanced theories including DFT+$U$, hybrid functionals, and many-body $G_0W_0$ approximation, using finite-temperature \textit{ab initio} MD simulations and Kubo-Greenwood (KG) formula (Supplemental Material \ref{sec:methods}). We find that ferrous iron at dodecahedral sites in pPv remains in the high-spin state up to 500 GPa (Supplemental Material \ref{sec:spin}). We establish a protocol to determine the appropriate mean-field starting point for $G_0W_0$ on highly correlated materials (Supplemental Material \ref{sec:starting}), and develop practical workflow for electronic conductivity calculations at the $GW$ level (Supplemental Material \ref{sec:methods}). We arrived at a convincing conclusion that electronic conduction plays only a minor role in pPv, marking an important step toward resolving the electrical conductivity in planetary deep interiors.

\begin{figure}%[!ht]
\includegraphics[width=0.48\textwidth]{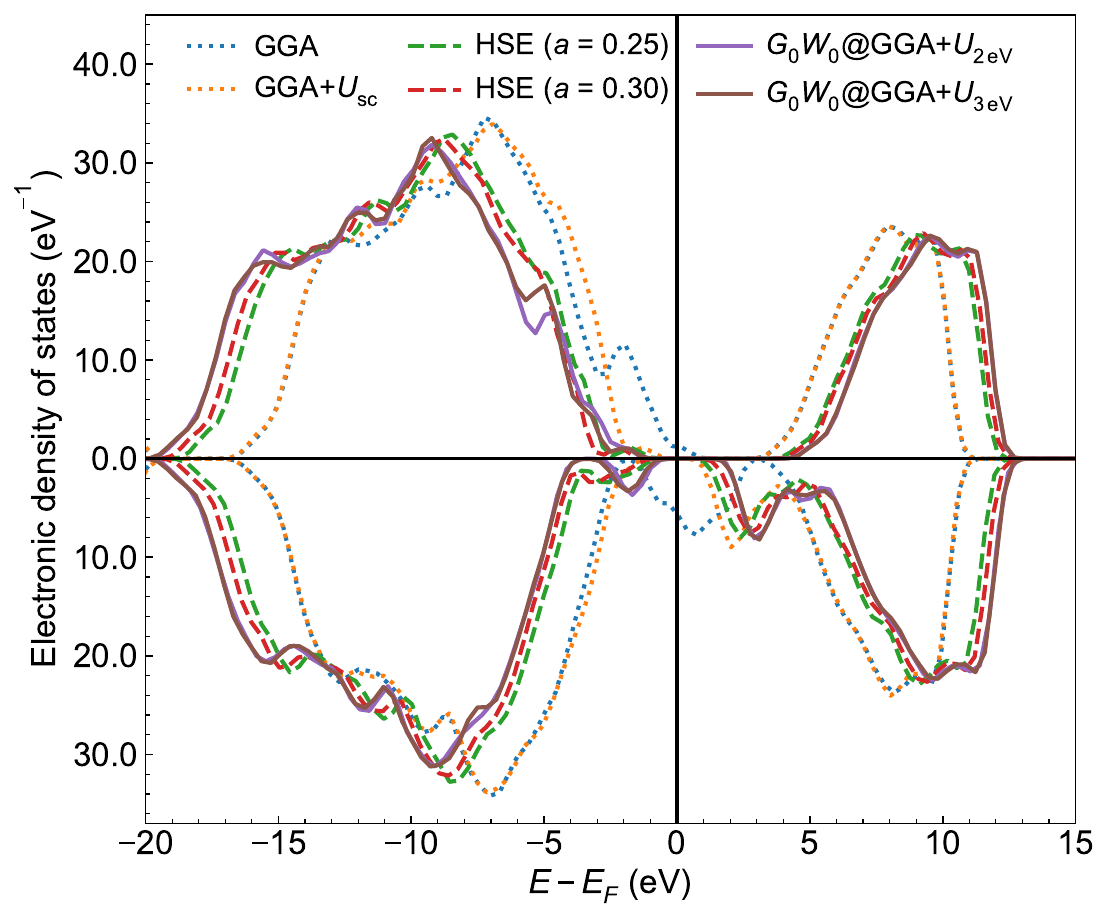}
\caption{Total electronic density of states of a molecular dynamics snapshot of (Mg$_{28}$Fe$_{4}$)Si$_{32}$O$_{96}$ pPv in high-spin state at 200~GPa and 4000~K obtained from different methods. The vertical black line represents the Fermi level ($E_F$). The upper and lower halves of the DOS plot correspond to the majority and minority spin states, respectively.}\label{fig:DOS}
\end{figure}

\begin{figure*}%[!ht]
\includegraphics[width=0.7\textwidth]{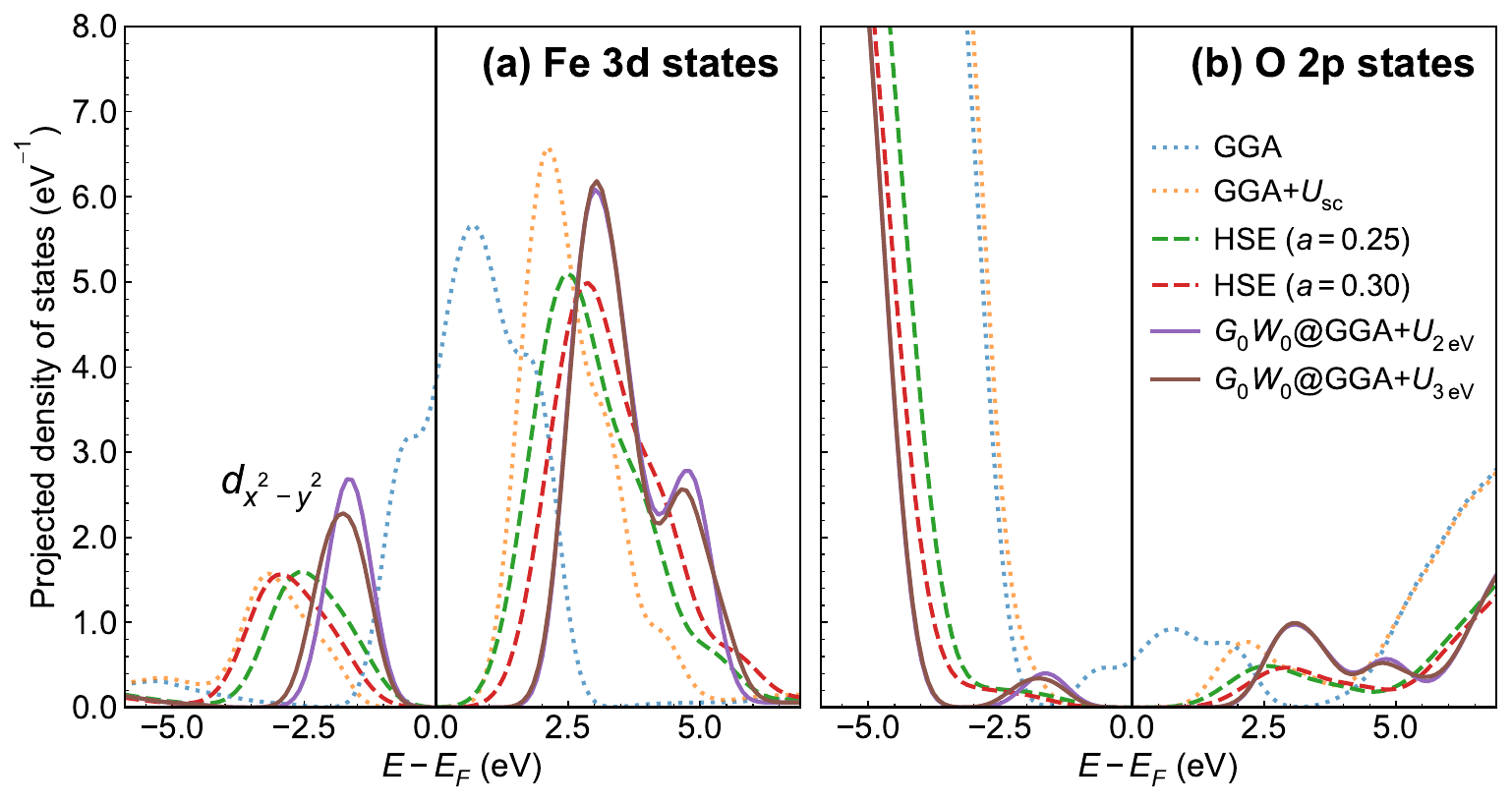}
\caption{Projected density of states (PDOS) for Fe-$3d$ states (a) and O-$2p$ states (b) in the same system as Fig.~\ref{fig:DOS} obtained from different methods. Only the minority spin channel is shown. The vertical black lines represent the Fermi level ($E_F$).}\label{fig:PDOS}
\end{figure*}

\textit{Electronic structure}---Accurately describing the electronic structure of pPv is a crucial prerequisite for obtaining reliable conductivity. Fig.~\ref{fig:DOS} and Fig.~\ref{fig:PDOS} present the total density of states (DOS) and projected density of states (PDOS) for (Mg$_{28}$Fe$_{4}$)Si$_{32}$O$_{96}$ pPv at 200~GPa and 4000~K, evaluated on the same molecular-dynamics snapshot with several electronic-structure levels: standard GGA, GGA+$U$ (with $U=2$ eV, $U=3$~eV and $U_{\rm sc}=5.26$ eV), HSE06 with Hartree-Fock (HF) exchange fraction $a=0.25$ and 0.30, and one-shot $G_0W_0$ starting from GGA and from GGA+$U$. The band structures of (Mg$_{3}$Fe)Si$_{4}$O$_{12}$ pPv at 0 K is also calculated (Fig.~\ref{fig:band}), showing consistently larger gaps due to the lack of thermal displacement of ions.

Standard GGA produces a spurious metallic solution with finite DOS at the Fermi level ($E_F$) (Fig.~\ref{fig:DOS}), mainly contributed by Fe-$3d$ states (Fig.~\ref{fig:PDOS}), consistent with over-delocalization from self-interaction error similar to FeO \cite{cococcioni_linear_2005}. Introducing a self-consistently computed Hubbard $U$ parameter (GGA+$U_{\rm sc}$), nonlocal exchange (HSE), or dynamical screening effect ($GW$) systematically shifts the majority-spin Fe-$3d$ states to lower energy and, in the minority-spin channel, splits the Fe-$3d$ manifold into $d_{x^{2}-y^{2}}$ states at the valence band and other $3d$ states ($d_{xy}$, $d_{xz}$, $d_{yz}$, and $d_{z^2}$) at the conduction band, thereby opening a well-defined gap around $E_F$. The on-site Hubbard $U$ correction primarily reorganizes Fe-$3d$ states but maintains similar O-$2p$ states compared to GGA, causing similar energy of O-$2p$ and Fe-$3d$ states and thus strong Fe--O hybridization at the valence band. On the contrary, HSE and $GW$ significantly shift O-$2p$ manifold away from the Fe-$3d$ edges by reducing the energy of O-$2p$ valence states, weakening $p$-$d$ hybridization, and nudging the system toward a more Mott-Hubbard-like regime \cite{fujimori_electronic_2001}. A similar functional dependence of the valence band character has been documented for olivine phosphates LiMPO$_4$ (M = Fe, Mn, Co, Ni): the application of Hubbard $U$ often pushes the transition-metal $3d$ manifold below the O-$2p$ states, yielding an O-$2p$-dominated valence band maximum (VBM), whereas HSE06, by placing O-$2p$ states at lower binding energy, restores a transition-metal-derived VBM required by the experimentally inferred structural stability during delithiation \cite{johannes_hole_2012,zhang_re-evaluation_2019}. By analogy, we regard the HSE and $GW$ electronic structures for Fe-bearing pPv as a more reliable description than GGA+$U{_\mathrm{sc}}$. Moreover, the two approaches---which both go beyond local Hubbard corrections by incorporating nonlocal exchange or dynamical screening---yield highly consistent DOS, further reinforcing the robustness of their accuracy.

Relative to the standard HSE06 ($a=0.25$), increasing HF exchange to $30\%$ slightly enlarges both the Fe-$3d$ splitting and the O-$2p$ separation, in line with established trends for transition-metal oxides under PBE0/Fock-0.5-type mixing \cite{tran_hybrid_2006}, producing a larger gap. To avoid inheriting an arbitrary parameter into the $G_0W_0$ results, we explicitly scanned GGA+$U$ starting points to achieve that the DOS is weakly dependent on the input $U$. For $U=0$--3 eV, the DOS of $G_0W_0$@GGA+$U$ is essentially invariant. However, GGA+$U$ itself with $U_{\mathrm{sc}}=5.26$ eV have already yielded a lower valence energy compared to $G_0W_0$@GGA+$U_{\mathrm{0-3\,eV}}$. Since $G_0W_0$ further shifts the valence energy downward, a potential overcorrection of the $d_{x^2-y^2}$ states by $G_0W_0$@GGA+$U_\mathrm{sc}$ is observed (Fig.~\ref{fig:PDOS-SI}). In addition, since the starting wave function with band energies similar to the $GW$ QP energies is considered better for $G_0W_0$ method \cite{fuchs_quasiparticle_2007, lany_band-structure_2013}, we regard $G_0W_0$@GGA+$U_{\mathrm{2-3\,eV}}$ as our most reliable QP reference and propose a similar starting-point sensitivity analysis as a general protocol for future $G_0W_0$ studies of transition-metal-bearing phases (see Supplemental Material \ref{sec:starting} for details).

The characteristics of DOS computed by configurations generated by MD simulations have direct implications for Kubo-Greenwood transport predictions. First, they demonstrate that thermal disorder at 4000~K does not restore metallicity once self-interaction is controlled: all beyond-GGA levels examined (GGA+$U_{\rm sc}$, HSE, $GW$) retain a clear band gap on the MD snapshot and suggest low intraband conductivity. Second, by widening the O-$2p$ separation and reducing $p$-$d$ hybridization at the band edges, HSE and $GW$ results show clear Mott-Hubbard character distinct from the mixed Mott-Hubbard/charge-transfer character obtained by GGA+$U_{\rm sc}$, suggesting a reduced $\sigma_{\mathrm{DC}}$. 

\begin{figure*}%[!ht]
\includegraphics[width=0.933\textwidth]{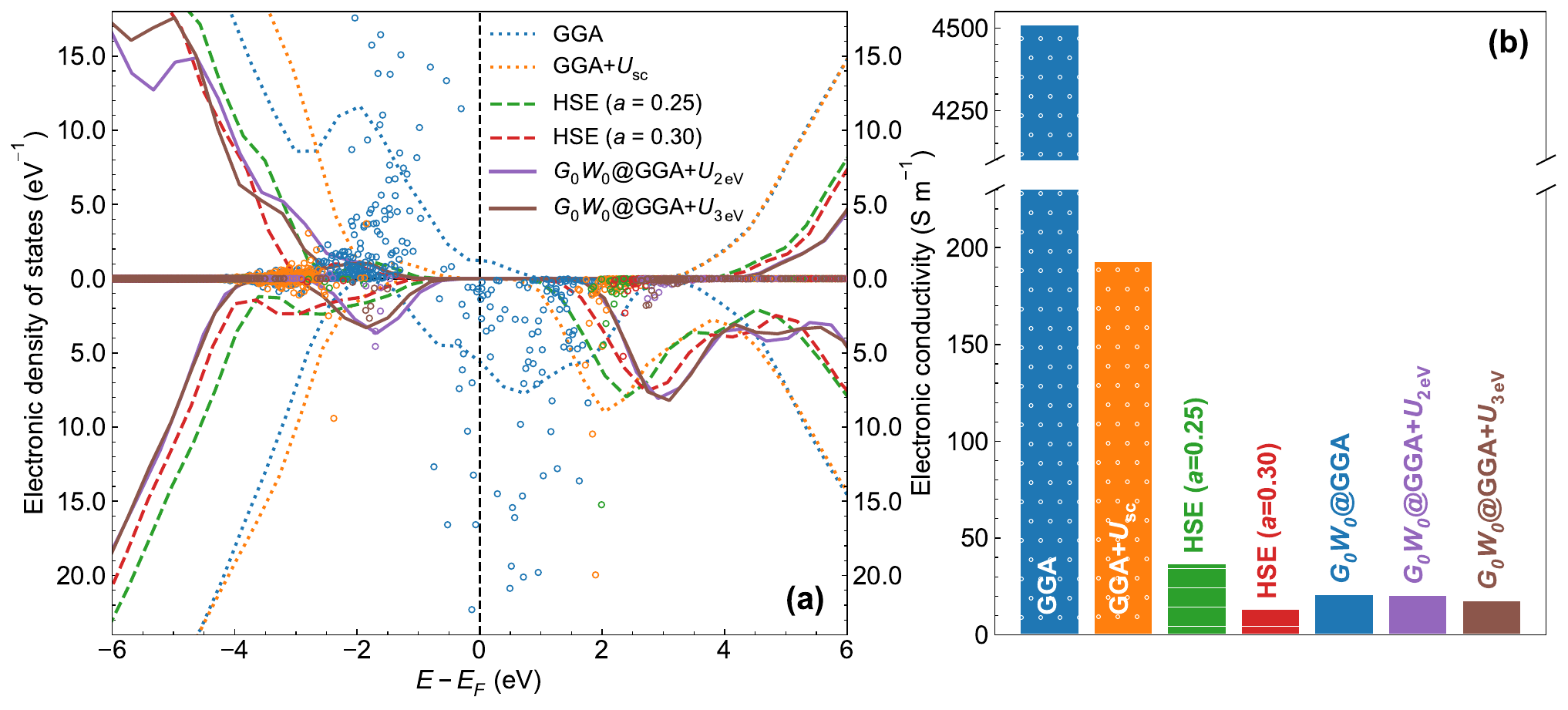}
\caption{(a) Decomposition of the Kubo-Greenwood DC conductivity of the same system as Fig.~\ref{fig:DOS} obtained from different methods. Each transition of electron is shown by a data point (at the average energy of the initial and final bands) that gives the contribution of that transition to the DC conductivity. The density of states in Fig.~\ref{fig:DOS} is also shown for comparison. The upward and downward directions of the plot represent the majority and minority spin states, respectively. (b) Total Kubo-Greenwood DC conductivity obtained from different methods, calculated as the summation of data points in (a).}\label{fig:decomp}
\end{figure*}

\textit{Electronic conductivity}---We evaluated the direct-current (DC) conductivity ($\sigma_{\mathrm{DC}}$) of the same system using the KG formula (Eq.~\eqref{eq:KG}), considering only the interband term because all levels of theory except standard GGA show a band gap. To resolve how specific bands contribute to the conductivity, we decomposed the KG conductivity into individual contributions from band-pair transitions:
\begin{equation}\label{eq:KG-decomp}
\begin{split}
\sigma_{m,m^{\prime}}(\omega) &=\frac{2 \pi e^2 \hbar^2}{3m_e^2 V \omega} \sum_{\alpha = 1}^3 \left(f\left(\epsilon_{m^{\prime}}\right) - f\left(\epsilon_m\right)\right) \\
& \times \left| \langle m | \nabla_\alpha | m^{\prime} \rangle \right|^2 \, \delta\!\left(\epsilon_{m} - \epsilon_{m^{\prime}}-\hbar \omega\right) ,
\end{split}
\end{equation}
The conductivity contributions are plotted in Fig.~\ref{fig:decomp}a on a common abscissa with the total DOS.

A pronounced spin asymmetry is observed (Fig.~\ref{fig:decomp}a and Fig.~\ref{fig:band}). In the majority-spin channel, all methods beyond standard GGA yield a wide gap, rendering this channel largely inert for low-energy electronic transport with small contribution to $\sigma_{\mathrm{DC}}$. However, since the two spin channels do not act as independent reservoirs and share a single chemical potential $E_F$, the majority-spin DOS is not symmetric with respect to $E_F$, causing thermally activated partial occupations near the valence band edge and a minor yet non-negligible contribution to $\sigma_{\mathrm{DC}}$ from the majority-spin channel, especially for GGA+$U_{\mathrm{sc}}$.

For GGA, the spurious metallicity produces an extensive manifold of partially filled states at $E_F$, leading to numerously large contributions (many exceeding the y-axis ranges of Fig.~\ref{fig:decomp}a) and an overestimated total $\sigma_{\mathrm{DC}}$ of $\sim\!4.5\times 10^3$ S\;m$^{-1}$. Introducing a self-consistent Hubbard correction (GGA+$U_{\mathrm{sc}}$) opens a gap and reduces $\sigma_{\mathrm{DC}}$ substantially; nevertheless, at the high temperature of 4000~K, the partial occupations at band edges still contribute to $\sigma_{\mathrm{DC}}$. After applying $U_{\mathrm{sc}}$, the Fe $d_{x^{2}-y^{2}}$ states become nearly degenerate with O-$2p$ states that are weakly affected by Hubbard $U$ at the valence band of the minority spin channel (Fig.~\ref{fig:band} and Fig.~\ref{fig:PDOS}). The resulting strong $p-d$ hybridization imparts a mixed Mott-Hubbard/charge-transfer character: both Fe-centered and ligand-centered holes appear, and a combination of $d\!\to\!d$, $p\!\to\!d$, and $p\!\to\!p$ transitions produces a total $\sigma_{\mathrm{DC}}$ of $\sim$200 S\;m$^{-1}$.

Hybrid DFT and the $GW$ approximation further reshape the pPv DOS and thus its electronic conductivity. The strongly reduced $p-d$ hybridization relative to GGA+$U_{\mathrm{sc}}$ confines thermally activated partial occupancies mainly to Fe-$3d$ states, causing $\sigma_{\mathrm{DC}}$ to drop below $\sim 40$ S\;m$^{-1}$. The four minority-spin $d_{x^{2}-y^{2}}$ states associated with the four Fe atoms form the valence band; thermally assisted transitions among these four states almost entirely control the valence-band contribution to the conductivity. Above $E_F$, the conduction band are dominated by the remaining 16 Fe-$3d$ states, with only a small admixture of O-$2p$ character arising from residual Fe-O hybridization; together these states govern the conduction-band contribution. HSE and $GW$ place the two Fe-$3d$-dominated DOS peaks at slightly different energies: in HSE they are systematically lower in energy than in $GW$, and thus HSE yields a larger conductivity contribution from the conduction band than from the valence band, whereas the opposite holds for $GW$. However, the resulting total conductivities remain comparable. The sharper band edges and deeper gap in the $GW$ DOS further deplete the near-edge electronic transport relative to HSE, pushing $\sigma_{\mathrm{DC}}$ down to $< 20$ S\;m$^{-1}$. Increasing the HF exchange fraction from 0.25 to 0.30 in HSE increases the gap, thereby lowering $\sigma_{\mathrm{DC}}$ and bringing it into better agreement with the $GW$ results. By contrast, within a range of $U$ values for the $G_0W_0$@GGA+$U$ ($U=0$--3 eV), the band-edge QP spectrum and $\sigma_{\mathrm{DC}}$ show no significant change, indicating desirable starting-point robustness at $GW$ level (Fig.~\ref{fig:decomp}b). Using a GGA+$U$ starting point with a larger $U$ value leads to potential overcorrection by $G_0W_0$ (Supplemental Material \ref{sec:starting}), but this would only increase the band gap and reduce $\sigma_{\mathrm{DC}}$, which does not affect our conclusions. Overall, $\sigma_{\mathrm{DC}}$ obtained from different levels of theory decreases systematically following $\mathrm{GGA} \gg \mathrm{GGA}+U_{\mathrm{sc}} > \mathrm{HSE}\;(a=0.25) > \mathrm{HSE}\;(a=0.30) \sim G_0W_0@\mathrm{GGA}+U$.
The progression reflects a systematic reduction of low-energy electronic transport as self-interaction is removed and dynamical screening is included.

\begin{figure}%[!ht]
\includegraphics[width=0.48\textwidth]{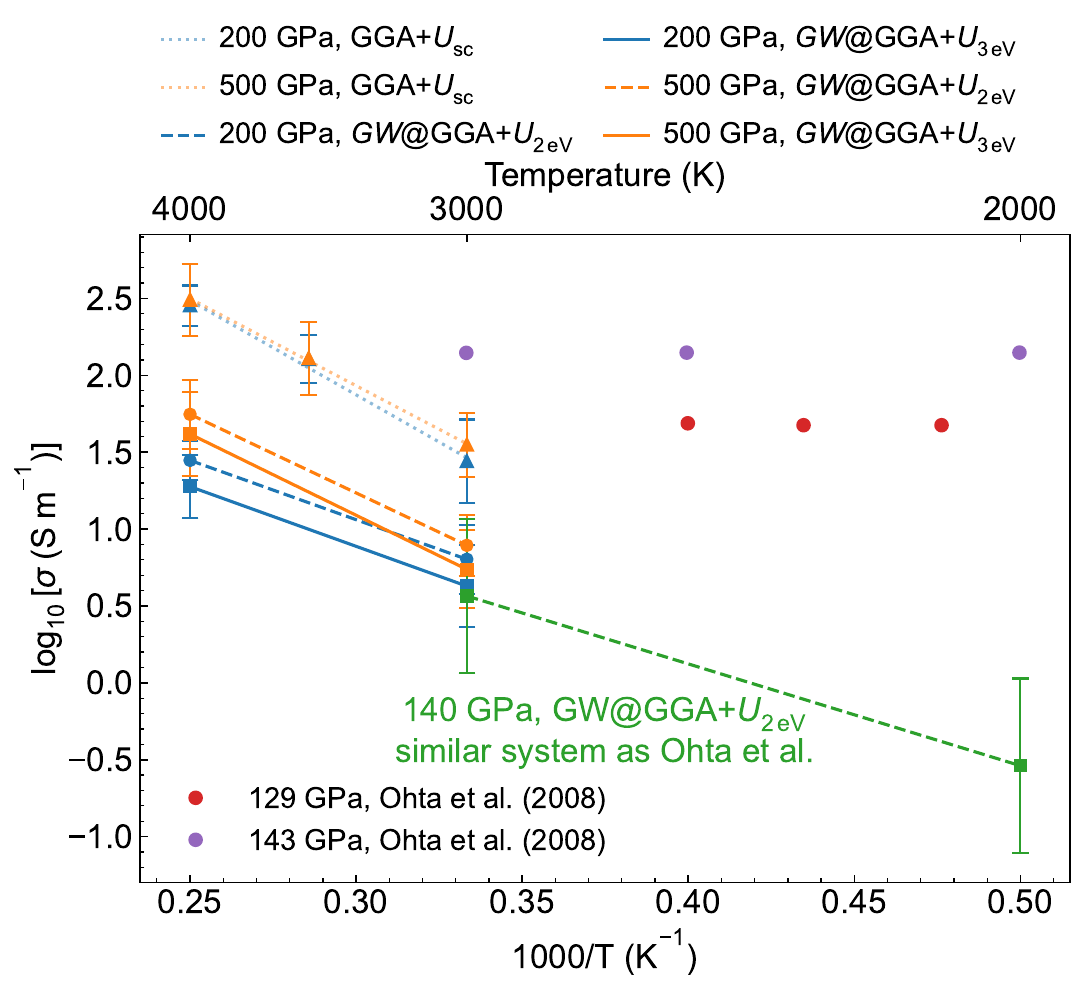}
\caption{Kubo-Greenwood DC conductivity of (Mg$_{28}$Fe$_{4}$)Si$_{32}$O$_{96}$ pPv (blue and orange) and (Mg$_{28}$Fe$_{4}$)(Si$_{31}$Al)O$_{96}$ pPv (green) in high-spin state under different pressure and temperature conditions. The experimental results of the bulk conductivity of (Mg$_{0.89}$Fe$_{0.11}$)SiO$_{3}$ pPv with Fe$^{3+}/\Sigma$Fe = 0.13 are shown for comparison \cite{ohta_electrical_2008}.}\label{fig:sigma}
\end{figure}

\textit{Electrical conductivity in planetary mantles}---According to previous experimental evidences, the electrical conductivity of Fe-bearing silicates under the Earth’s mantle condition is mostly dominated by small polaron conduction, involving hopping of charged polarons between Fe\textsuperscript{2+} and Fe\textsuperscript{3+} sites \cite{yoshino_laboratory_2010, yoshino_electrical_2013, yoshino_electrical_2016}. In contrast to small polaron conduction characterized by localized charge transport, band conduction (i.e., electronic conduction), which is the only conduction mechanism captured by the KG formula in this study, involves delocalized electrons thermally excited into extended electronic states. The band conduction mechanism was proposed to become dominant in (Mg,Fe)O and overtake small polaron conduction at temperatures above $\sim$2000~K \cite{holmstrom_electronic_2018}. To evaluate whether a similar crossover occurs in Fe-bearing pPv, we calculated $GW$-level $\sigma_{\mathrm{DC}}$ over a range of pressures and temperatures and compared them with experimental measurements under the Earth's core-mantle boundary conditions \cite{ohta_electrical_2008}. We performed MD simulations and the KG conductivities averaged over 10 independent snapshots are reported (Fig.~\ref{fig:sigma}).

The structure we have discussed contains only ferrous iron occupying dodecahedral (A) site. Nevertheless, in natural systems and experimental samples, ferric iron can be incorporated into mantle silicates via charge-coupled substitutions \cite{huang_effect_2021}, with potential impacts on electronic structure and transport. For example, in the only available conductivity experiment on Fe-bearing pPv to date, the (Mg$_{0.89}$Fe$_{0.11}$)SiO$_{3}$ sample had Fe$^{3+}/\Sigma$Fe $= 0.13 \pm 0.10$ \cite{ohta_electrical_2008}; in Al,Fe-bearing pPv, Fe$^{3+}/\Sigma$Fe has been reported to exceed 50\%, commonly attributed to Fe$^{3+}$--Al$^{3+}$ coupled substitution \cite{sinmyo_ferric_2006, sinmyo_ferric_2008}. To facilitate a direct comparison with the experiment, we built a (Mg$_{28}$Fe$_{4}$)(Si$_{31}$Al)O$_{96}$ pPv supercell that approximates the sample's total Fe content and Fe$^{3+}/\Sigma$Fe as closely as practicable. The Fe$^{3+}$ is introduced via Fe$^{3+}$--Al$^{3+}$ coupled substitution to obtain Fe$^{3+}/\Sigma$Fe = 0.25 (about twice the experimental fraction owing to supercell size constraints). Since the Fe$^{3+}$ still occupies the A site, the HS state remains energetically favorable up to 200 GPa \cite{yu_spin_2012}. Using $G_0W_0$@GGA+$U$, we evaluated its conductivity considering HS Fe at 140 GPa. Our results suggest that, under comparable temperature--pressure--composition conditions, the electronic conductivity of Fe-bearing pPv remains $< 5\%$ of the experimentally measured total conductivity (Fig.~\ref{fig:sigma}). At 3000 K, the (Mg$_{28}$Fe$_{4}$)(Si$_{31}$Al)O$_{96}$ pPv yields $\sim 3.6$ S\;m$^{-1}$ at 140 GPa, which is close to and slightly lower than the value for (Mg$_{28}$Fe$_{4}$)Si$_{32}$O$_{96}$ pPv at 200 GPa, indicating that the presence of ferric iron does not substantially enhance electronic conduction. Ferric substitution at the level considered here does not collapse the gap nor increase $p-d$ hybridization within the $GW$ framework. Due to the limited ionic diffusivity \cite{ammann_first-principles_2010}, the ionic conductivity in pPv is likely negligible. Therefore, under Earth’s core-mantle boundary conditions, small polaron hopping probably contributes to over 95\% of the electrical conductivity of pPv.

The pressure range in which pPv can stably exist within super-Earth mantles may extend up to 500~GPa \cite{umemoto_phase_2017, dutta_high-pressure_2023}. For (Mg$_{0.875}$Fe$_{0.125}$)SiO$_{3}$ pPv at 200--500~GPa and 3000--4000~K, $\sigma_{\mathrm{DC}}$ computed by both GGA+$U_{\mathrm{sc}}$ and $G_0W_0$@GGA+$U$ exhibits weak pressure dependence but a clear thermal activation, with Arrhenius fits yielding activation enthalpies of 1.5--2 eV; however, GGA+$U_{\mathrm{sc}}$ systematically overestimates the conductivity by nearly an order of magnitude (Fig.~\ref{fig:sigma}). Even at 500~GPa and 4000~K---conditions corresponding to the core-mantle boundary of a 4-Earth-mass super-Earth with an Earth-like potential temperature---the electronic conductivity predicted by $G_0W_0$@GGA+$U$ does not exceed 60 S\;m$^{-1}$. We find that pPv with Earth-like iron contents is unlikely to metallize in super-Earth mantles via band closure; small-polaron conduction may remain the dominant mechanism under the conditions we examined. In contrast, iron-rich (Mg,Fe)O and FeO have reported metallization under Earth's lower mantle conditions \cite{ohta_experimental_2012, ohta_highly_2014}. Our GW-level results show that such metallization is not universal among Fe-bearing mantle phases: the dominant silicate phase pPv retains a robust gap, emphasizing strong mineralogical control on deep-mantle conductivity. However, early planetary evolution may involve extremely high interior temperatures and partial melting due to accretionary heating \cite[e.g.,][]{stixrude_melting_2014, vazan_contribution_2018}. Additionally, exoplanetary mantles may have diverse chemical compositions \cite{hatalova_compositional_2025}, potentially hosting significantly more Fe-rich silicates. Whether metallization occurs in pPv under these conditions requires further investigation, and a full assessment of the electrical conductivity of Fe-bearing pPv will require explicit polaron hopping models beyond the band-to-band KG treatment.

Finally, this study underscores a methodological point: KG-based conductivity predictions are only as accurate as the underlying electronic structure, and care must be taken to employ \textit{ab initio} techniques that faithfully capture the intrinsic electronic properties of the target materials.

\textit{Acknowledgments}---This work was funded by the National Science Foundation under Grant EAR-2242946 to J.D. The work by Y.Z. and X.W. was supported by Seven Research, LLC. The work by S.Z. was supported by the Department of Energy (National Nuclear Security Administration), University of Rochester, National Inertial Confinement Program, under Award No. DE-NA0004144. The authors are pleased to acknowledge that the work reported in this paper was substantially performed using Princeton University’s Research Computing resources. This research used resources of the National Energy Research Scientific Computing Center, a DOE Office of Science User Facility supported by the Office of Science of the U.S. Department of Energy under Contract No. DE-AC02-05CH11231 using NERSC award BES-ERCAP0035544. This work used Anvil supercomputer at Purdue University through allocation EES250058 from the Advanced Cyberinfrastructure Coordination Ecosystem: Services \& Support (ACCESS) program, which is supported by U.S. National Science Foundation grants \#2138259, \#2138286, \#2138307, \#2137603, and \#2138296.

\textit{Data availability}---The data that support the findings of
this article are openly available \cite{peng_ab_2026}.


\bibliographystyle{apsrev4-2}
\bibliography{refs}

\clearpage

\renewcommand{\thefigure}{S\arabic{figure}}
\setcounter{figure}{0}
\setcounter{page}{1}
\renewcommand{\thetable}{S\arabic{table}}
\renewcommand{\appendixname}{}

\appendix

\onecolumngrid

\begin{titlepage}
\begin{center}

% \vspace*{1cm}

{\Large Supplemental Material for}

\vspace{0.5cm}

{\Large\bfseries Ab initio electronic conductivity of Fe-bearing post-perovskite}

\vspace{0.5cm}

{Yihang Peng\,\orcidlink{0000-0003-4404-8973},$^1$ Yupei Zhang\,\orcidlink{0009-0000-5835-4619},$^2$ Shuai Zhang\,\orcidlink{0000-0001-9503-4964},$^3$ Chenxing Luo\,\orcidlink{0000-0003-4116-6851},$^{1,4}$\\ Donghao Zheng\,\orcidlink{0009-0003-0497-4961},$^1$ Nelson Naveas\,\orcidlink{0000-0003-0949-277X},$^5$ Xifan Wu\,\orcidlink{0000-0001-6561-8942},$^2$ and Jie Deng\,\orcidlink{0000-0001-5441-2797}$^1$}

\begin{center}
    \textit{
    $^1$Department of Geosciences, Princeton University, Princeton, NJ, USA\\
    $^2$Department of Physics, Temple University, Philadelphia, PA, USA\\
    $^3$Laboratory for Laser Energetics, University of Rochester, Rochester, NY, USA\\
    $^4$Department of Applied Physics and Applied Mathematics, Columbia University, New York, NY, USA\\
    $^5$Departamento de Física Aplicada and Instituto Universitario de Ciencia de Materiales “Nicolás Cabrera” (INC),\\
    Universidad Autonoma de Madrid, Campus de Cantoblanco, Madrid, Spain
    }
\end{center}

\vfill

\end{center}
\end{titlepage}

\twocolumngrid

\section{Methods}\label{sec:methods}
\subsection{GGA+\textit{U}}
GGA+$U$ calculations were performed with Quantum ESPRESSO (QE) \cite{giannozzi_quantum_2009, giannozzi_advanced_2017} code in PBEsol approximation \cite{perdew_restoring_2008}. We utilized the projector-augmented wave (PAW) method with PAW datasets from the PSlibrary \cite{dal_corso_pseudopotentials_2014}. The valence electronic configurations are 3s$^2$3p$^6$3d$^6$4s$^2$, 2s$^2$2p$^6$3s$^2$, 3s$^2$3p$^2$, and 2s$^2$2p$^4$ for Fe, Mg, Si, and O, respectively. We chose an energy cutoff of 80 Ry for electronic wave functions and 500 Ry for spin-charge density and potentials. The convergence threshold of all self-consistent field calculations was $1 \times 10^{-6}$ Ry. The Hubbard correction \cite{anisimov_band_1991} using the simplified formulation of Ref.~\cite{dudarev_electron-energy-loss_1998} was applied to Fe-3d states with orthogonalized atomic orbitals. The Hubbard parameter $U$ was either set to fixed values of 2 eV and 3 eV, or self-consistently computed based on DFPT \cite{timrov_hubbard_2018} using the HP \cite{timrov_hp_2022} code implemented in QE following the approach described in previous studies on iron oxides \cite{naveas_first-principles_2023, naveas_first-principles_2023_2, naveas_dft_2025}. The convergence threshold for the response function was $1 \times 10^{-6}$ Ry along with a q-point mesh of $2 \times 2 \times 2$. The self-consistent calculations of the Hubbard $U$ parameter ($U_\mathrm{sc}$) were performed through an iterative procedure, involving structural optimization followed by recalculation of the Hubbard parameters for each newly relaxed crystal structure \cite{hsu_spin-state_2011, timrov_self-consistent_2021}. The $U_\mathrm{sc}$ values were iterated until its difference between two successive iterations was smaller than 0.01 eV, ensuring self-consistency. The converged $U_\mathrm{sc}$ values under different spin states and pressure conditions are reported in Table.~\ref{tab:U}. The 20-atom unit cell of C\textit{mcm} MgSiO$_3$ pPv with one Mg\textsuperscript{2+} substituted by Fe\textsuperscript{2+} at a dodecahedral (A) site was used for the band structure calculations, while a $2 \times 1 \times 1$ supercell with one Mg\textsuperscript{2+} substituted by Fe\textsuperscript{2+} was used for all other static calculations. The Brillouin zone of the 20-atom unit cell, 40-atom supercell, and 160-atom supercell (Section \ref{sec:sigma}) were sampled by $10 \times 3 \times 4$, $5 \times 3 \times 4$, and $2 \times 3 \times 2$ Monkhorst-Pack meshes, respectively. Structure optimizations were performed by relaxing atom positions with convergence thresholds of 0.005~eV/\AA{} for the atomic force, 0.5 kbar for the pressure, and $1 \times 10^{-6}$ Ry for the total energy. Phonon calculations were performed using the finite-displacement method with the PHONOPY \cite{togo_first_2015} code and GGA+$U_{\rm sc}$ forces calculated from QE. The vibrational contributions to the free energy were calculated using the quasiharmonic approximation \cite{wallace_thermodynamics_1972} implemented in the \textit{qha} \cite{qin_qha_2019} code.

\subsection{Hybrid functionals}
\textit{Ab initio} calculations using the HSE06 hybrid functional \cite{heyd_hybrid_2003, paier_screened_2006} were performed using the PAW dataset with an energy of 800~eV as implemented in the Vienna Ab Initio Simulation Package (VASP) \cite{kresse_efficient_1996}. The valence electronic configurations are consistent with GGA+$U$ calculations. The HSE06 functional is a screened implementation of the PBE0 functional. The exchange-correlation energy of PBE0 is given by
\begin{equation}\label{eq:hse}
    E_{\rm xc}^{\rm PBE0} = a E_{\rm x}^{\rm HF} + (1-a) E_{\rm x}^{\rm PBE} + E_{\rm c}^{\rm PBE}
\end{equation}
where the exchange energy contains both Hartree-Fock (HF) ($E_{\rm x}^{\rm HF}$) and PBE \cite{perdew_generalized_1996} ($E_{\rm x}^{\rm PBE}$) exchange terms, whereas the correlation energy is from the PBE functional. The HSE06 functional partitions the exchange term into short-range and long-range components using a screening parameter optimized to balance computational efficiency and accuracy. The parameter $a$ denotes the HF exchange mixing coefficient, i.e., the fraction of HF exchange. The value of $a$ is 0.25 in standard PBE0 and HSE06 functionals derived from perturbation theory \cite{perdew_rationale_1996}. However, it is suggested that the optimal $a$ value that reproduces the experimental data is actually system-dependent \cite{heyd_efficient_2004, meng_when_2016}. In this study, we employed both the standard HSE06 functional with $a = 0.25$ and a modified version with a higher mixing parameter of $a = 0.30$. The latter yields results that show improved agreement with those obtained from many-body $GW$ approximation. The $8 \times 2 \times 3$, $3 \times 2 \times 3$, and $1 \times 1 \times 1$ $k$-point grids were used for the 20-atom unit cell, 40-atom supercell, and 160-atom supercell, respectively, to reduce the computational cost. To ensure a direct comparison of electronic structure results obtained from different methods, the HSE calculations are directly performed on the configurations obtained from GGA+$U_{\rm sc}$ structure optimizations and MD simulations.

\subsection{\textit{GW} approximation}
The one-shot $G_0W_0$ calculation for the QP energy from many-body perturbation theory was performed with the BerkeleyGW package \cite{hybertsen_electron_1986, deslippe_berkeleygw_2012} and QE, starting from both GGA and GGA+$U$ wave functions, for the 160-atom MD snapshots described in Section \ref{sec:sigma}. We utilize the multiple-projector norm-conserving pseudopotentials generated by the Optimized Norm-Conserving Vanderbilt PSeudopotential package \cite{hamann_optimized_2013} with consistent valence electronic configurations. A single $\Gamma$ point in the Brillouin zone was used to maintain a reasonable computational cost. The static inverse dielectric matrix is constructed within the random phase approximation (RPA), and the Hybertsen-Louie generalized plasmon-pole (HL-GPP) model is used to extend the dielectric response to non-zero frequencies. We employ a screened cutoff of 10 Ry and 10000 bands to ensure convergence of the QP gap at the $G_{0}W_{0}$ level.

\subsection{Electronic conductivity}\label{sec:sigma}
The electronic density of states and electronic conductivity were calculated from ionic trajectories obtained by \textit{ab initio} MD simulations using the GGA+$U_{\rm sc}$ functional. \textit{Ab initio} MD simulations were performed by VASP under the canonical ensemble with the Nosé-Hoover thermostat \cite{hoover_canonical_1985}, sampling the Brillouin zone at the $\Gamma$ point with energy cutoffs of 500~eV and an energy tolerance of $10^{-4}$~eV. Finite temperature effects on the energy were considered using the Mermin functional with the Fermi-Dirac smearing \cite{mermin_thermal_1965, wentzcovitch_energy_1992}. The initial configurations for the MD simulations were generated by substituting 12.5 mol\% Mg with Fe in the $4 \times 1 \times 2$ supercell of C\textit{mcm} MgSiO$_3$ pPv with 160 atoms, corresponding to an iron content similar to the Earth's mantle. The spatial distribution of Fe atoms was determined using the special quasi-random structure method implemented in the Alloy-Theoretic Automated Toolkit \cite{jiang_first-principles_2004, van_de_walle_efficient_2013}, which minimizes an objective function for each supercell to homogenize the local atomic environment. For each pressure and temperature condition, we extracted 10 snapshots from the equilibrated MD trajectory of at least 5 picoseconds in length, ensuring that they were sufficiently separated in time to be uncorrelated.

\begin{figure*}%[h!]
\includegraphics[width=0.6\textwidth]{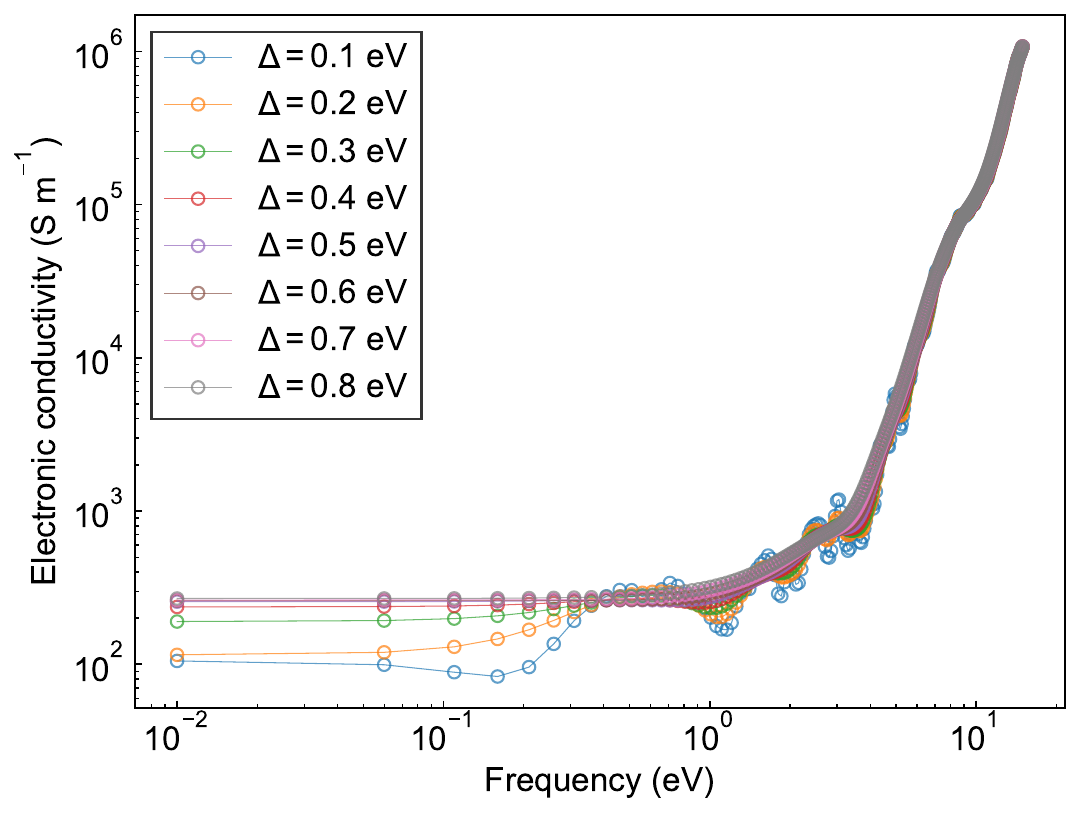}
\caption{Kubo-Greenwood conductivity of (Mg$_{28}$Fe$_{4}$)Si$_{32}$O$_{96}$ pPv with high-spin iron as a function of frequency at 4000~K and 500~GPa calculated using GGA+$U_{\rm sc}$. The conductivities of one identical molecular dynamics snapshot calculated using different broadening parameters $\Delta$ are shown for comparison.}\label{fig:Delta}
\end{figure*}

\begin{figure*}%[h!]
\includegraphics[width=0.6\textwidth]{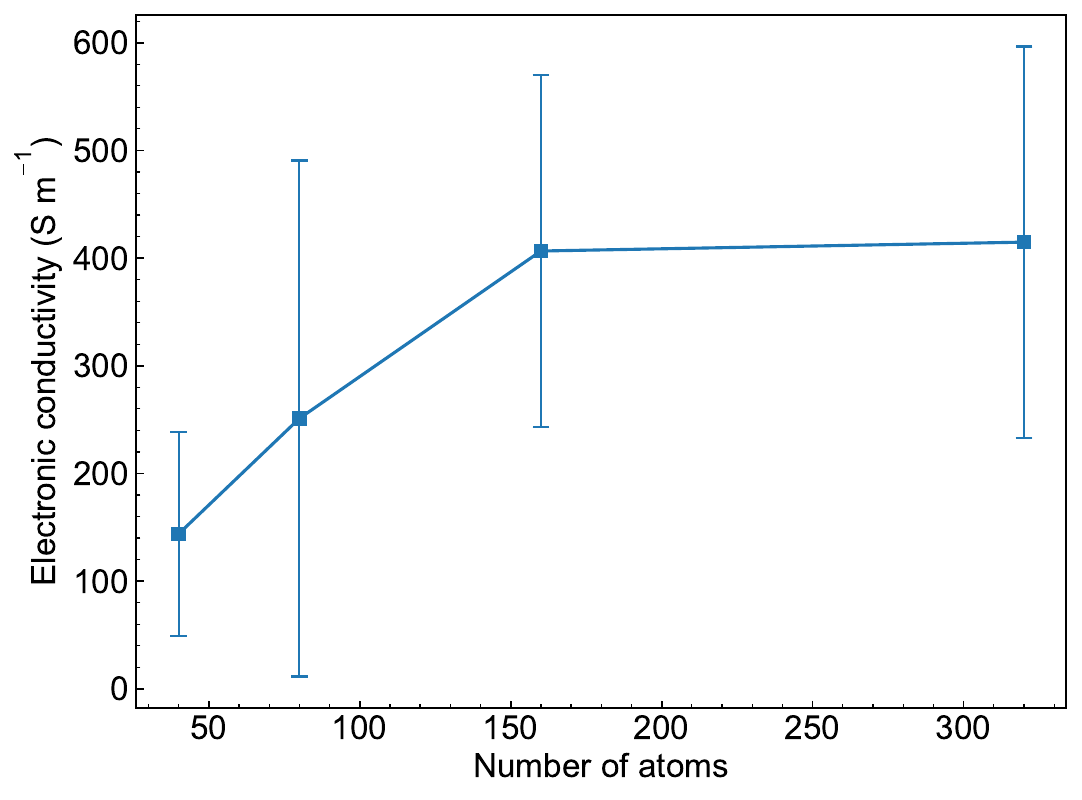}
\caption{Kubo-Greenwood DC conductivity of (Mg$_{28}$Fe$_{4}$)Si$_{32}$O$_{96}$ pPv in high-spin state as a function of the number of atoms in the supercell at 4000~K and 500~GPa calculated using GGA+$U_{\rm sc}$. The error bars represent the standard deviation of electrical  conductivity obtained from different molecular dynamics snapshots.}\label{fig:size}
\end{figure*}

The electronic conductivity $\sigma$ of each snapshot was calculated using the Kubo-Greenwood (KG) formula \cite{kubo_statistical-mechanical_1957, greenwood_boltzmann_1958} as implemented in the KGEC \cite{calderin_kubogreenwood_2017} code for QE and kgoptics \cite{desjarlais_electrical_2002} code for VASP:
\begin{equation}\label{eq:KG}
\begin{split}
\sigma(\omega) &=\frac{2 \pi e^2 \hbar^2}{3m_e^2 V \omega} \sum_{\alpha = 1}^3 \sum_{m,m^{\prime}} \left(f\left(\epsilon_{m^{\prime}}\right) - f\left(\epsilon_m\right)\right) \\
& \times \left| \langle m | \nabla_{\alpha} | m^{\prime} \rangle \right|^2 \, \delta\!\left(\epsilon_{m} - \epsilon_{m^{\prime}}-\hbar \omega\right) ,
\end{split}
\end{equation}
where $\omega$ is the frequency of the electric current; $\hbar$ is the reduced Planck constant; $\delta$ is the Dirac $\delta$ function; $e$ and $m_e$ are the electron charge and mass; the $\alpha$ sum runs over the three spatial directions; $m$ and $m'$ denote the Kohn-Sham (KS) wave functions with corresponding eigenvalues $\epsilon_m$, $\epsilon_{m'}$, and associated Fermi-Dirac occupation numbers $f(\epsilon_m)$ and $f(\epsilon_{m'})$. This approach has been applied for various planet-forming materials under extreme conditions \cite{scipioni_electrical_2017, soubiran_electrical_2018, holmstrom_electronic_2018, stixrude_silicate_2020, ghosh_melting_2024}. The $\delta$ function in Eq.~\eqref{eq:KG} is approximated by a Gaussian with a broadening parameter $\Delta$ of 0.5~eV following Ref.~\cite{ghosh_melting_2024}, and the effect of $\Delta$ on the DC conductivity results (Fig.~\ref{fig:Delta}) is comparable with the statistical uncertainty of the conductivity calculated from different MD snapshots (Fig.~\ref{fig:size}). Since the Eq.~\eqref{eq:KG} becomes unphysical when $\hbar \omega < \Delta$, the direct-current (DC) conductivity $\sigma_\mathrm{DC}$ is obtained by linearly extrapolating to zero frequency. Notably, as the only open-source implementation of KG calculation compatible with QE, KGEC does not support the post-processing of spin-polarized DFT calculations \cite{calderin_kubogreenwood_2017}. We therefore modified the original implementation to enable conductivity calculations for spin-polarized systems by adding a sum over spin states in Eq.~\eqref{eq:KG}. The finite-size effect on the electrical conductivity is evaluated by conducting MD simulations and KG conductivity calculations on supercells of different sizes using the GGA+$U_{\rm sc}$ functional. We find that a supercell containing 160 atoms used in this study provides sufficiently converged results (Fig.~\ref{fig:size}).

\begin{figure*}%[h!]
\includegraphics[width=0.6\textwidth]{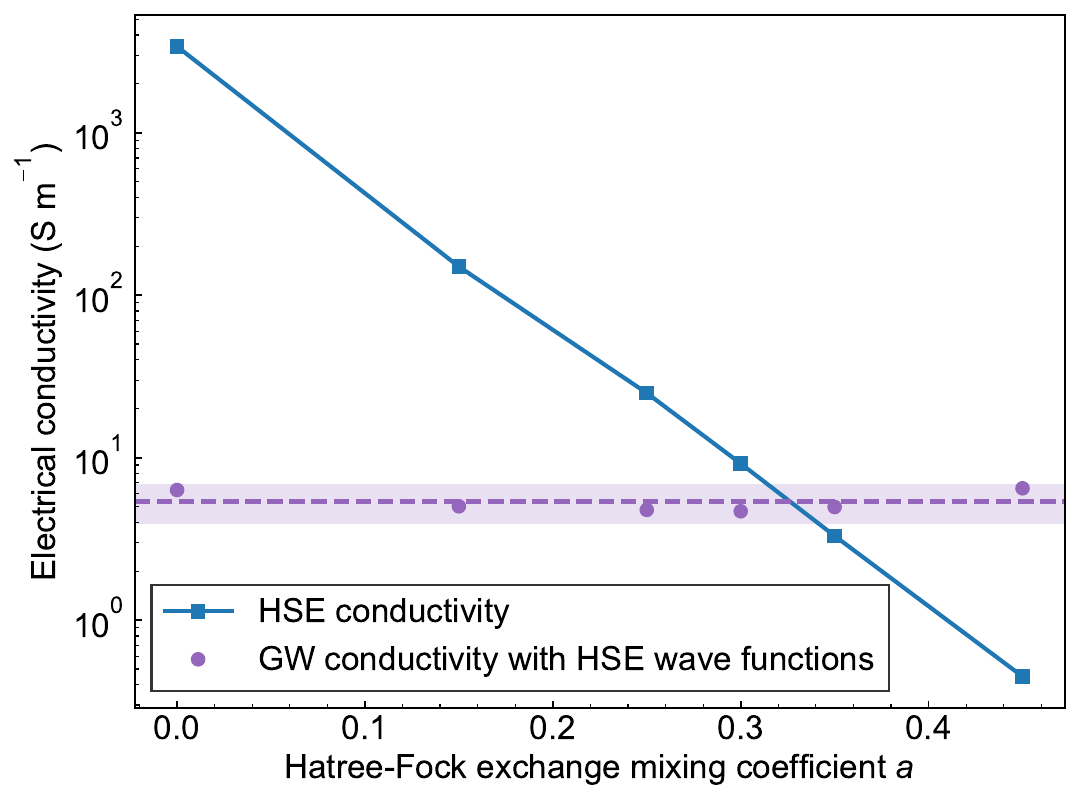}
\caption{Total Kubo-Greenwood DC conductivity of (Mg$_{28}$Fe$_{4}$)Si$_{32}$O$_{96}$ pPv in high-spin state at 4000~K and 500~GPa. Blue squares: conductivity results from HSE functional as a function of fraction of Hartree-Fock exchange $a$ (for $a = 0$, the functional becomes PBE). Purple circles: The Kohn-Sham wave functions are identical to those of the blue squares, but eigenvalues are replaced by $GW$ quasiparticle energies calculated from a GGA starting point. The purple dashed line shows the mean $GW$ conductivity, and the shaded area indicates the standard deviation.}\label{fig:benchmark}
\end{figure*}

In the $G_0W_0$ approximation used in this study, we apply state-resolved corrections to the KS eigenvalues while keeping the KS eigenfunctions unchanged (i.e., a diagonal self-energy in the KS basis). Accordingly, in our KG conductivity calculations, we evaluate the velocity matrix elements $\langle m| \nabla_{\alpha} |m^{\prime}\rangle$ with the starting-point wave functions (e.g., from DFT or DFT+$U$), while using the QP energies and corresponding Fermi-Dirac occupation numbers at the target temperature. Our tests show that, once the band energies are fixed by the QP corrections, replacing the KS wave functions by those from different functionals produces no appreciable change in the DC conductivity (Fig.~\ref{fig:benchmark}). This insensitivity suggests that, for the systems studied, the DC conductivity is primarily governed by the energy spectrum rather than modest variations in the wave functions, thereby justifying the use of starting-point matrix elements combined with $G_0W_0$-corrected energies.

\section{Spin state of iron in post-perovskite}\label{sec:spin}

\begin{figure*}%[!ht]
\includegraphics[width=\textwidth]{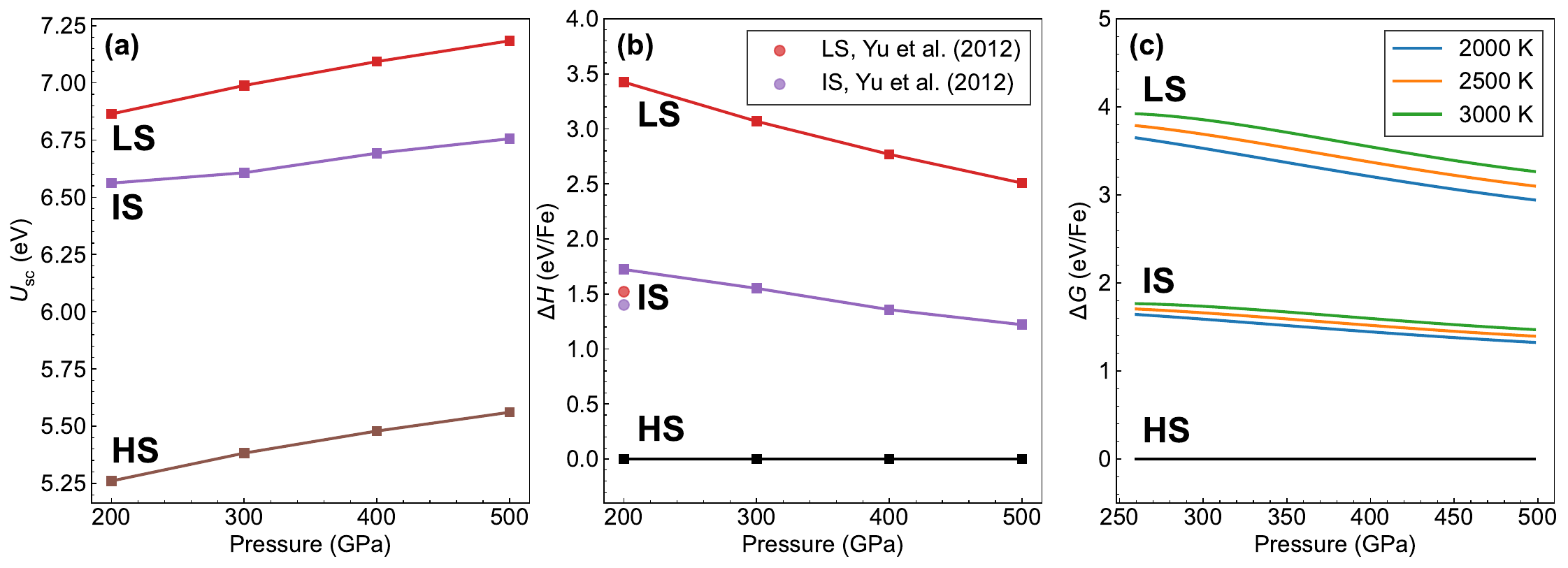}
\caption{Self-consistent Hubbard $U$ value (a),  enthalpy (b), and Gibbs energy (c) of (Mg$_{7}$Fe)Si$_{8}$O$_{24}$ pPv in high-spin (HS), intermediate-spin (IS), and low-spin (LS) states. Previous DFT+$U_{\rm sc}$ results \cite{yu_spin_2012} are shown for comparison. Both enthalpy and Gibbs energy values are normalized to those of HS states. } \label{fig:spin}
\end{figure*}

In mantle phases, iron may experience a pressure-induced reduction in total electron spin ($S$), a phenomenon known as spin-state crossover, which can profoundly influence the physical and chemical properties of the host minerals \cite{badro_spin_2014}, including their conductivity \cite{holmstrom_electronic_2018, stixrude_silicate_2020}. Previous enthalpy calculations using DFT+$U_{\rm sc}$ have shown that below 200~GPa, ferrous iron occupying A site in pPv remains in the high-spin (HS, $S = 2$) state, while intermediate-spin (IS, $S = 1$) and low-spin (LS, $S = 0$) states are highly unfavorable \cite{yu_spin_2012}. However, the higher pressure conditions, as well as the potential effect of high temperature in super-Earth mantles on iron spin state in pPv, remain unexplored. 

To address this issue, we first calculate the self-consistent Hubbard $U$ parameters of HS, IS, and LS pPv under super-Earth mantle pressures, which are used for the following GGA+$U_{\rm sc}$ calculations (Fig.~\ref{fig:spin}a and Table.~\ref{tab:U}). The trends of slightly increasing $U_{\rm sc}$ with pressure and decreasing $U_{\rm sc}$ with increasing total electron spin are generally consistent with previous studies \cite{hsu_spin_2010, hsu_spin-state_2011, yu_spin_2012, sun_ldausc_2020}. Fig.~\ref{fig:spin}b and Fig.~\ref{fig:spin}c present the relative enthalpy and relative Gibbs energies of pPv with different spin states, under pressures of 200--500~GPa. The electronic entropy ($S_{\mathrm{el}}$), vibrational entropy ($S_{\mathrm{vib}}$), and magnetic entropy ($S_{\mathrm{mag}} = k_B \sum_i \ln (\mu_i + 1)$ \cite{holmstrom_spin_2015, grimvall_spin_1989} where $\mu_i$ is the total magnetic moment) are considered when calculating the Gibbs energy. Our calculated enthalpy difference between the IS and HS states at 200~GPa has excellent agreement with the previous study \cite{yu_spin_2012}. The discrepancy in the LS state may arise from the fact that, compared to the PBE functional and ultrasoft pseudopotentials used in Ref.~\cite{yu_spin_2012}, we employed the GGA functional and PAW datasets, as the thermophysical properties of GGA align very well with experiments \cite[e.g.,][]{deng_melting_2023}. Our results show that even under pressures as high as 500~GPa, the HS state of pPv remains significantly lower in enthalpy compared to the IS and LS states ($\Delta H_\mathrm{LS} - \Delta H_\mathrm{HS} \approx 2.5$ eV/Fe at 500 GPa). Moreover, the relatively higher $S_{\mathrm{vib}}$ and $S_{\mathrm{mag}}$ associated with the HS state further stabilize it under the high-temperature conditions of planetary interiors. Considering potential further phase transitions in pPv at $\sim$500~GPa \cite{umemoto_phase_2017, dutta_high-pressure_2023}, we conclude that (Mg$_{0.875}$Fe$_{0.125}$)SiO$_3$ pPv remains in the HS state under super-Earth mantle conditions, with no reduction in spin moment expected. Therefore, the HS state is the only one considered in the subsequent calculations.

\section{Band structure and charge density of Fe-bearing post-perovskite}\label{sec:band}

\begin{figure*}%[!ht]
\includegraphics[width=\textwidth]{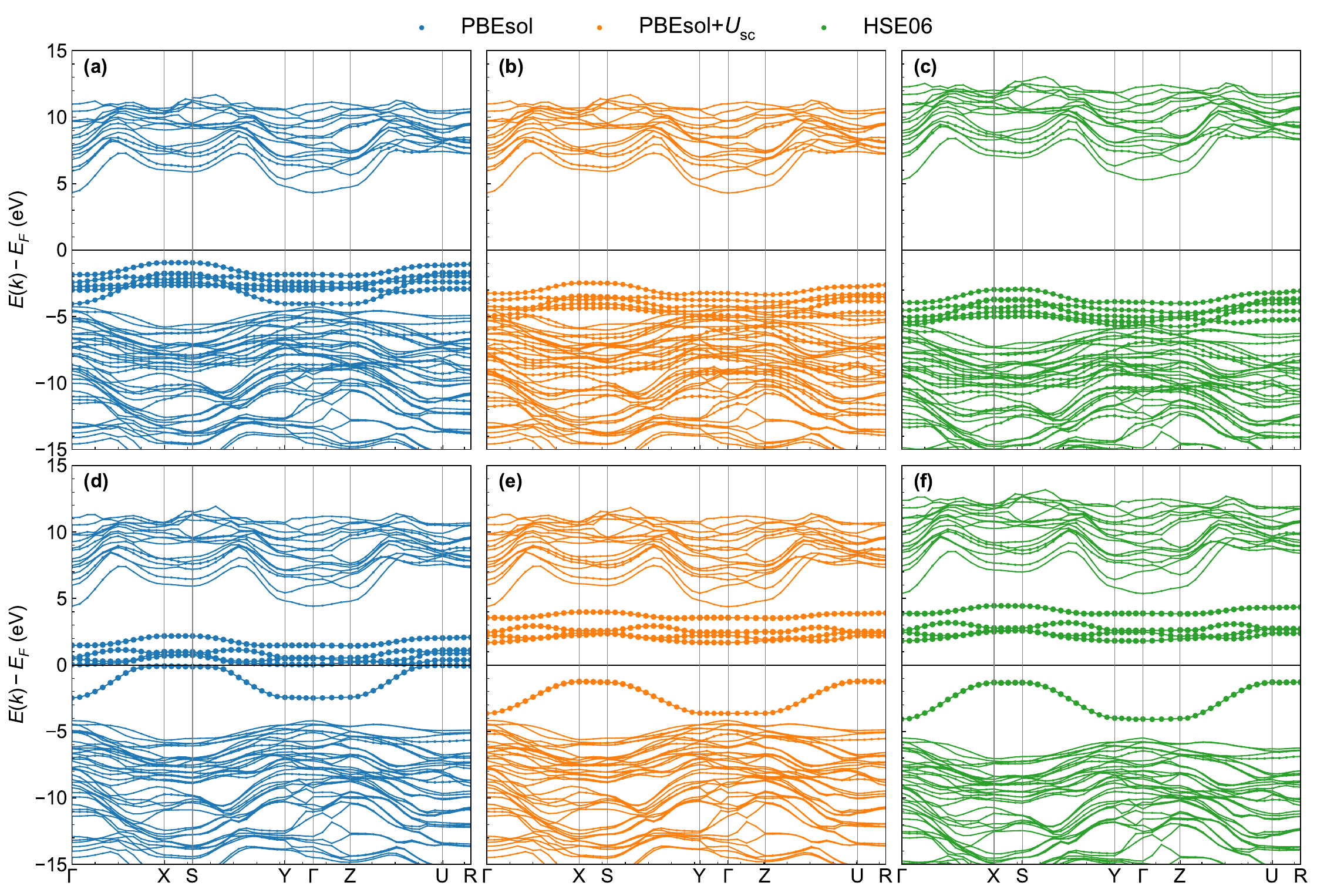}
\caption{Electronic band structure of (Mg$_{3}$Fe)Si$_{4}$O$_{12}$ pPv in high-spin state at 200~GPa computed with GGA (a, d), GGA+$U_{\rm sc}$ (b, e), and HSE06 (c, f) functionals. The symbol size encodes the weight of Fe-$3d$ states. (a, b, c) Majority spin; (d, e, f) Minority spin. Because the calculation of the band structure requires using the pPv unit cell, introducing a single Fe atom results in an iron concentration that is twice that of the other systems reported in this study.}\label{fig:band}
\end{figure*}

Fig.~\ref{fig:band} presents spin-resolved band structures of HS (Mg$_{3}$Fe)Si$_{4}$O$_{12}$ pPv at 200~GPa computed with GGA, GGA+$U_{\rm sc}$, and standard HSE06 with $a = 0.25$. Across all functionals, the five bands closest to the Fermi level ($E_F$) carry dominant Fe-$3d$ character, indicating that low-energy excitations are controlled by the Fe manifold and its hybridization with O-$2p$ states (see also Fig.~\ref{fig:PDOS}). In the majority channel (Fig.~\ref{fig:band}a,b,c), all functionals yield a wide gap ($\gtrsim$5.2~eV from valence-band maximum to conduction-band minimum). In the minority channel (Fig.~\ref{fig:band}d,e,f), standard GGA produces a metallic solution with Fe-$3d$ bands crossing $E_F$. Introducing on-site correlation via GGA+$U_{\rm sc}$ splits the minority $d$ manifold by making four $d$ bands shift upward and one shifts downward, opening a gap of $\sim$3~eV. HSE06 yields a very similar ordering and a slightly larger separation of Fe-$3d$ bands compared to GGA+$U_{\rm sc}$, which reflects similarly reduced self-interaction, but additionally shifts other bands: non-Fe states in the deeper valence and conduction regions move to lower and higher energies relative to GGA, respectively. By contrast, GGA+$U_{\rm sc}$ shows almost identical band structure with GGA except for Fe-$3d$ states.

\begin{figure*}%[!ht]
\includegraphics[width=0.7\textwidth]{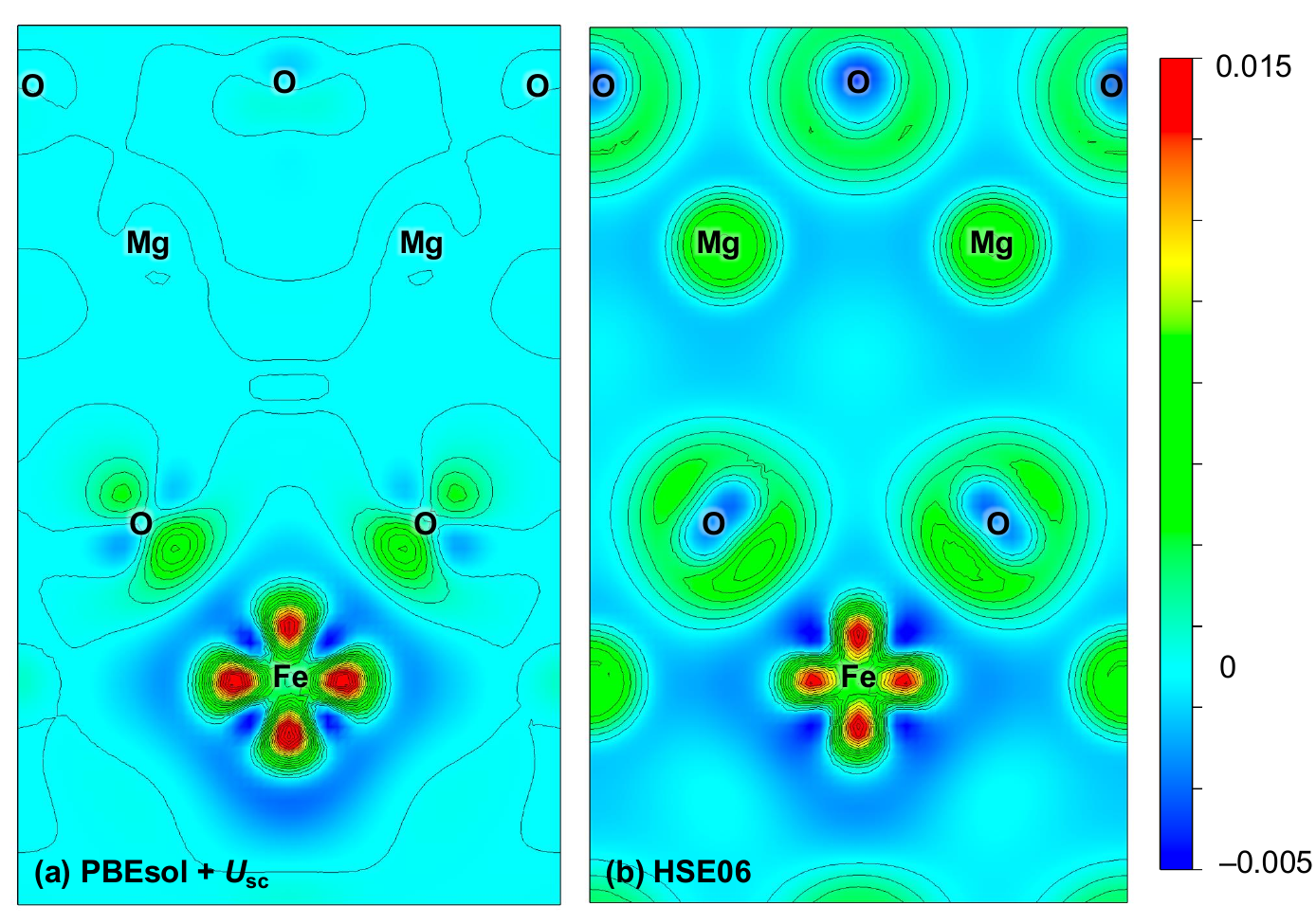}
\caption{Contour plots of the charge density ($\rho$) difference between different functionals on the (001) plane of (Mg$_{7}$Fe)Si$_{8}$O$_{24}$ pPv in high-spin state at 200~GPa. Red, yellow, and green indicate an accumulation of charge relative to GGA; blue indicates depletion. (a) $\rho_{\mathrm{GGA}+U_{\rm sc}}-\rho_{\mathrm{GGA}}$; (b) $\rho_{\mathrm{HSE06}}-\rho_{\mathrm{GGA}}$.}\label{fig:charge}
\end{figure*}

Fig.~\ref{fig:charge} presents charge-density differences on the (001) plane of (Mg$_{7}$Fe)Si$_{8}$O$_{24}$ pPv intersecting the Mg and Fe sites, computed as $\rho_{\mathrm{GGA}+U_{\rm sc}}-\rho_{\mathrm{GGA}}$ (Fig.~\ref{fig:charge}a) and $\rho_{\mathrm{HSE06}}-\rho_{\mathrm{GGA}}$ (Fig.~\ref{fig:charge}b). Around Fe, both maps exhibit a four-lobe motif aligned with the $x$ and $y$ directions, i.e., the characteristic footprint of a $d_{x^2-y^2}$ orbital on this crystallographic cut. This anisotropy is expected because the majority-spin Fe-$3d$ manifold is fully occupied (thus largely isotropic in its net contribution) whereas, in the minority channel, only the lowest-energy $d_{x^2-y^2}$ state is occupied and contributes to the charge density, preserving its in-plane anisotropy. This finding is consistent with previous DFT+$U_{\rm sc}$ results \cite{yu_spin_2012}.

The charge density difference caused by Hubbard $U$ correction is highly localized: only Fe and the first-shell O ligands display noticeable changes, while more distant O and Mg sites are nearly unaffected (Fig.~\ref{fig:charge}a). By correcting self-interaction on Fe-$3d$ manifolds and generating a more localized Fe-centered pattern, Hubbard $U$ also slightly reduces spurious Fe-O $p$-$d$ hybridization in GGA and induces a compensating, oxygen-centered rebonding, shown as the reduced contribution of O-$2p$ states to the DOS at $E_F$ (Fig.~\ref{fig:PDOS}b). HSE06 produces a similar Fe-centered pattern but with a lattice-wide response, i.e., all Mg and O atoms show charge accumulation in the near-valence region. This nonlocal redistribution reflects the short-range exact-exchange component of HSE06, which reduces self-interaction for all valence states, compacts cation charge, and pushes O-$2p$ bands away from the Fe-$3d$ manifold (deepening the O-$2p$ valence and raising O-$2p$ contributions in the conduction band, see Fig.~\ref{fig:band} and Fig.~\ref{fig:PDOS}). A clear trend is observed from GGA to GGA+U and further to hybrid functionals, where the first-shell O ligands of Fe exhibit increasingly ionic character, aligning with earlier studies on hematite ($\rm \alpha-Fe_2O_3$) \cite{liao_testing_2011}.

These real-space charge redistribution reinforces the interpretation drawn from the band structure. First, the tightening of Fe-$3d$ charge and reduced $p$ admixture at the band edges imply smaller conductivity under GGA+$U_{\rm sc}$ and HSE06. Second, the contrast between the highly local response under GGA+$U_{\rm sc}$ and the system-wide redistribution under HSE06 clarifies why HSE06 yields a larger gap and stronger suppression of low-energy conductivity: the hybrid functional does not simply localize Fe-$3d$ electrons; it re-partitions charge across the entire lattice, systematically reducing both the self-interaction and the $p$-$d$ hybridization that would enhance near-edge transitions.

\section{Starting point dependence and protocol for \textit{G}\textsubscript{0}\textit{W}\textsubscript{0} calculations}\label{sec:starting}

\begin{figure*}%[!ht]
\includegraphics[width=0.7\textwidth]{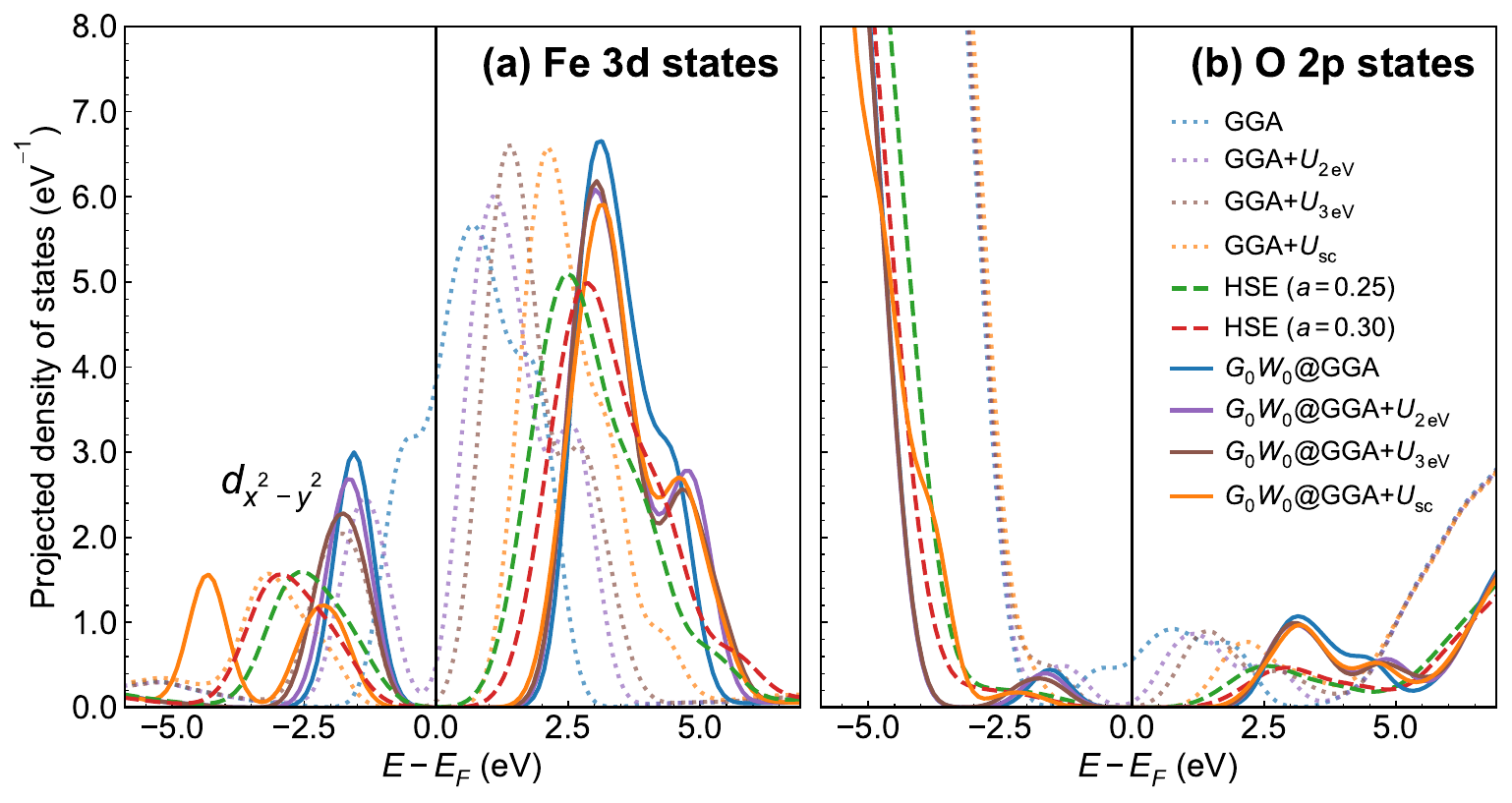}
\caption{Projected electronic density of states (PDOS) for Fe and O atoms in the same system as Fig.~\ref{fig:DOS} obtained from different methods. (a) Fe-$3d$ states; (b) O-$2p$ states. Only the minority spin channel is shown. The vertical black lines represent the Fermi level ($E_F$).The semi-transparent dotted curves, shown in the same color as the solid curves for $GW$ quasiparticle PDOS, represent the PDOS of their DFT starting points.}\label{fig:PDOS-SI}
\end{figure*}

A central aim of this study is to establish a $G_0W_0$ protocol whose QP energies are weakly dependent on the mean-field starting point, so that after a reasonable choice of input wave functions the $G_0W_0$ correction converges to a stable result rather than inheriting an arbitrary, system-specific parameter. Otherwise, as in DFT+$U$ or hybrid-functional calculations, an uncontrolled degree of freedom (i.e., the input $U$ and $a$) would preclude definitive conductivity predictions.

Because the on-site $U$ acts primarily on Fe-$3d$ states, the O-$2p$ PDOS is essentially insensitive to the $G_0W_0$ starting point and remains close to the HSE spectrum with $a=0.30$ (Fig.~\ref{fig:PDOS-SI}b). By contrast, a clear starting-point dependence appears in the Fe sector (Fig.~\ref{fig:PDOS-SI}a). For starting values $U=0$, 2, and 3 eV, the QP DOS is remarkably consistent: the Fe-derived peaks lie near $E_F-1.7$ eV (valence band, dominated by $d_{x^2-y^2}$) and $E_F+3.5$ eV (conduction band, dominated by other Fe-$3d$ states), i.e., the desired regime of starting-point insensitivity. However, when the starting U is increased to the self-consistent value $U_{\mathrm{sc}}=5.26$ eV, the $d_{x^2-y^2}$ states at the valence-band maximum undergo an additional splitting, evolving from a single peak at $\sim E_F-1.7$ eV into two lower-energy peaks. This change also feeds back onto the ligand manifold via enhanced hybridization, producing a discernible modification of the O-$2p$ DOS (Fig.~\ref{fig:PDOS-SI}b).

We interpret the discrepancy between $G_0W_0$@GGA+$U_{\mathrm{sc}}$ and the other $G_0W_0$ results as an overcorrection driven by an overlocalized starting point. For $U=0$ or 2 eV, the mean-field $d_{x^2-y^2}$ valence peak sits above the $GW$ QP energy; at $U=3$ eV, the pre- and post-$GW$ positions are nearly identical, with $GW$ imparting only a slight upward shift toward---evidence that the QP result has effectively converged with respect to input $U$. In general, because $GW$ includes dynamical screening, it tends to increase the band gap from the mean-field starting point \cite{rodl_quasiparticle_2009, liao_testing_2011, lany_band-structure_2013}, realized as a net downward shift of valence energies and an upward shift of conduction energies. If the starting $U$ is too large, reduced screening in the large-gap starting point further amplifies the self-energy, yielding a spurious extra splitting and an unphysically deep valence edge. Consistent with this picture, the conduction-band QP PDOS is nearly independent of the starting $U$ (including $U_{\mathrm{sc}}$), because all starting points underestimate the conduction energies in a similar way that $GW$ can robustly correct.

Because one-shot $G_0W_0$ is a perturbative correction to a mean-field Hamiltonian, one seeks a starting functional whose KS band energies already approximate the QP energies \cite{fuchs_quasiparticle_2007, lany_band-structure_2013}. Prior work on iron oxides has shown that $G_0W_0$ based on DFT+$U$ yields dielectric function and photoemission spectra in best agreement with experiment \cite{liao_testing_2011}. Although GGA+$U_{\mathrm{sc}}$ provides a respectable local description of Fe-$3d$ physics, i.e., predicting a $3d$ gap close to the $GW$ result and widely used in prior studies of Fe spin states in similar systems \cite{hsu_spin_2010, hsu_spin-state_2011, yu_spin_2012}, it is not an optimal starting point for $G_0W_0$ as it simultaneously underestimates the energies of valence band and the conduction band. Therefore, the $G_0W_0$@GGA+$U_{\mathrm{2\,eV}}$ and $G_0W_0$@GGA+$U_{\mathrm{3\,eV}}$ results reported here probably provide the most reliable electronic structure among the levels of theory we tested. In future studies of other materials, a similar protocol can be adopted: determine an appropriate $G_0W_0$ starting point by analyzing how the key features of the KS DOS and QP DOS evolve as a function of the input parameters.

% \section{Supplementary figures and tables}

\begin{table*}
\caption{Self-consistent Hubbard $U$ value of (Mg$_{7}$Fe)Si$_{8}$O$_{24}$ pPv in high-spin (HS), intermediate-spin (IS), and low-spin (LS) states under different pressures.}\label{tab:U}
\begin{tabular}{ccccc}
\toprule
   & 200 GPa & 300 GPa & 400 GPa & 500 GPa \\ \midrule
HS & 5.26    & 5.38    & 5.48    & 5.56    \\
IS & 6.56    & 6.61    & 6.69    & 6.76    \\
LS & 6.86    & 6.99    & 7.09    & 7.18    \\ \bottomrule
\end{tabular}
\end{table*}

\end{document}